\documentclass[12pt]{article}
\usepackage{graphicx}

\usepackage{scicite}

\usepackage{times}

\newenvironment{sciabstract}{%
\begin{quote} \bf}
{\end{quote}}

\newcounter{lastnote}
\newenvironment{scilastnote}{%
\setcounter{lastnote}{\value{enumiv}}%
\addtocounter{lastnote}{+1}%
\begin{list}%
{\arabic{lastnote}.}
{\setlength{\leftmargin}{.22in}}
{\setlength{\labelsep}{.5em}}}
{\end{list}}

\title{The Formation of Massive Star Systems by Accretion}

\author
{Mark R. Krumholz,$^{1\ast}$ Richard I. Klein,$^{2,3}$ Christopher F. McKee,$^{2,4}$ \\
Stella S.~R. Offner,$^{4}$ and Andrew J. Cunningham$^{3}$\\
\\
\normalsize{$^{1}$Department of Astronomy, University of California, Santa Cruz}\\
\normalsize{201 Interdisciplinary Sciences Building, Santa Cruz, CA 95064, USA}\\
\normalsize{$^{2}$Department of Astronomy, University of California, Berkeley}\\
\normalsize{601 Campbell Hall, Berkeley, CA 94720-3411, USA}\\
\normalsize{$^{3}$Lawrence Livermore National Laboratory, AX Division}\\
\normalsize{7000 East Avenue, Livermore, CA 94550, USA}\\
\normalsize{$^{4}$Department of Physics, University of California, Berkeley}\\
\normalsize{366 LeConte Hall, Berkeley, CA 94720-7300, USA}\\
\normalsize{$^\ast$To whom correspondence should be addressed. E-mail:  krumholz@ucolick.org.}
}

\date{}

\newcommand{\msun}{M_{\odot}}
\newcommand{\lsun}{L_{\odot}}

\newcommand{\vecx}{\mathbf{x}}
\newcommand{\vecv}{\mathbf{v}}
\newcommand{\vecp}{\mathbf{p}}
\newcommand{\dt}{\frac{\partial}{\partial t}}
\newcommand{\kp}{\kappa_{\rm 0P}}
\newcommand{\kr}{\kappa_{\rm 0R}}
\newcommand{\trk}{T_{\rm r}}

\newcommand{\mnras}{Mon.\ Not.\ R.\ Astron.\ Soc.}
\newcommand{\aj}{Astronom.\ J.}
\newcommand{\apj}{Astrophys.\ J.}
\newcommand{\apjl}{Astrophys.\ J.\ Lett.}
\newcommand{\nat}{Nature}
\newcommand{\aap}{Astron.\ \& Astrophys.}
\newcommand{\jcp}{J.\ Comp.\ Phys.}
\newcommand{\araa}{Ann.\ Rev.\ Astron.\ \& Astrophys.}

\begin{document} 


\baselineskip12pt


\maketitle


\begin{sciabstract}
Massive stars produce so much light that the radiation pressure they exert on the gas and dust around them is stronger than their gravitational attraction, a condition that has long been expected to prevent them from growing by accretion. We present three-dimensional radiation-hydrodynamic simulations of the collapse of a massive prestellar core and find that radiation pressure does not halt accretion. Instead, gravitational and Rayleigh-Taylor instabilities channel gas onto the star system through non-axisymmetric disks and filaments that self-shield against radiation, while allowing radiation to escape through optically-thin bubbles. Gravitational instabilities cause the disk to fragment and form a massive companion to the primary star. Radiation pressure does not limit stellar masses, but the instabilities that allow accretion to continue lead to small multiple systems.
\end{sciabstract}


Stars can form with masses up to at least $120$ times that of the Sun \cite{weidner04,figer05}, but the mechanism by which the most massive stars form is a longstanding mystery. Stars with masses greater than $\sim 20$ $\msun$ have Kelvin times (the time required for a star to radiate away its gravitational binding energy) that are shorter than their formation times, and as a result they attain their full luminosities while still accreting from their natal clouds. As the radiation from such an embedded, massive star diffuses outward through the dusty gas in the protostellar envelope, it exerts a force that opposes gravity. Spherically averaged, the ratio of the radiative and gravitational forces is $7.7\times 10^{-5} \kappa_0 (L/M)_0$, where $\kappa_0$ is the specific opacity of the gas in cm$^2/\mbox{g}$ and $(L/M)_0$ is the stellar light-to-mass ratio in units of $\lsun/\msun$. Because the dusty envelopes of massive protostars have $\kappa_0\sim 5$, the ratio of radiative to gravitational force exceeds unity for all stars with $(L/M)_0 \geq 2500$. Main sequence stars reach this value of light-to-mass ratio at masses $\sim 20$ $\msun$. Therefore, radiation is expected to halt spherically symmetric infall \cite{kahn74,wolfire87} near this mass. Two-dimensional simulations of massive star formation, which include rotation and thus an accretion disk that partially shields the gas from radiation \cite{nakano95,jijina96} find that radiation completely halts accretion once stars reach $\sim 40$ $\msun$ \cite{yorke02}, inconsistent with the highest known stellar masses.

Here we report a three-dimensional simulation of the formation of stars with masses greater than $20$ $\msun$ that includes the effects of radiation pressure. Three-dimensionality is important because instabilities that determine the interaction of gas and radiation, such as Rayleigh-Taylor instabilities, show faster growth rates and higher saturation amplitudes in three dimensions \cite{marinak95a}. Additionally, instabilities in accretion disks \cite{shu90,krumholz07a}, and the resulting formation of companion stars \cite{kratter08a}, can only be simulated when the disk is represented non-axisymmetrically. It is important to consider these effects because most massive stars are members of multiple systems \cite{mason98a,sana08a}.

Our initial conditions consisted of a gas cloud with mass $=100$ $\msun$, radius $=0.1$ pc, and density profile $\rho\propto r^{-1.5}$, consistent with models \cite{mckee02,mckee03} and observations \cite{beuther07c} of the initial states of massive prestellar cores. Its initial temperature was 20 K and it was in slow, solid-body rotation at a rate such that the ratio of rotational kinetic energy to gravitational binding energy was $0.02$, which is consistent with the rotation rates seen in lower-mass cores \cite{goodman93}. Previous two-dimensional simulations suggest that varying these parameters within the observed range of massive core properties would not alter the qualitative behavior \cite{yorke02}. Even though observed massive cores have large turbulent velocities \cite{beuther07c}, we did not include turbulence in our initial conditions in order to focus on the effects of radiation pressure. Simulations that include the effects of turbulence \cite{krumholz07a} do not appear to produce qualitatively different results.

We evolved this initial state using our adaptive mesh refinement code ORION \cite{klein99,krumholz04,fisher02,krumholz07b,shestakov08}, which solves the equations of gravito-radiation-hydrodynamics in the gray, flux-limited diffusion approximation. The code dynamically increases resolution as needed down to a minimum cell size of 10 AU; regions that collapse to densities above the Jeans density at the maximum resolution become star particles, each of which produces a luminosity determined by a protostellar evolution model (see SOM for details).

The simulation passed through several distinct phases (Fig.\ 1, S1--S3, and Movie S1), forming multiple stars (Fig.\ 2). The cloud began to collapse immediately and a central protostar formed 3.6 kyr afterward. For next next 17 kyr, the protostar accreted smoothly via an axisymmetric disk (Fig.\ 1A). During this phase the mass of the star grew to 11 $\msun$, and its luminosity stayed below $\sim 10^4$ $\lsun$. Because $(L/M)_0<1000$, radiation pressure produced no noticeable effects. After $\sim 20$ kyr the disk became gravitationally unstable and developed a pronounced two-armed spiral that transported angular momentum efficiently. (Fig.\ 1B) \cite{lodato05}. Accretion onto the protostar continued smoothly. 

Accretion, unimpeded by radiation pressure, continued until $25-26$ kyr, when the star reached a mass of roughly 17 $\msun$ and achieved a sustained light to mass ratio $(L/M)_0 \approx 2500$, driven by Kelvin-Helmholtz contraction. This was luminous enough for its radiation pressure force to exceed the gravitational force, and the star began to drive gas outward around the polar axis, inflating radiation-filled bubbles both above and below the accretion disk (Fig.\ 1C). The density inside the bubbles was very low, and within them the radiation pressure exceeded the gas pressure by orders of magnitude. Almost all gas falling onto the protostar struck the walls of the bubbles, where it was shocked and swept up into the bubble walls. However, this did not slow accretion in our simulation, because the gas that struck the bubble walls eventually traveled along the margin until it reached the disk, at which point it continued to accrete onto the star.  During this phase, radiation forces acting on material accreting onto the disk and gravitational forces acting on material in the disk caused it to become increasingly non-axisymmetric. A series of small secondary stars formed in the disk, most of which advected inward due to dynamical friction with the gas and collided with the central protostar. As a result the accretion rate onto the central star became variable, but its mean value remained roughly unchanged.

Around 35 kyr a series of disk-borne stars collided and became massive enough to resist being dragged inward (Fig.\ 1D). The secondary star was initially smaller than the primary and orbited it, intercepting and accreting much of the inflowing gas \cite{bate00}. As a result the secondary star acquired its own disk and grew faster at first than the primary, and the system reached a mass ratio $>0.5$. Thereafter, the disk continued to fragment but at a much-diminished rate, and accretion became almost evenly divided between the two massive stars. A third small disk-borne star was ejected into a wide orbit in our simulation, but eventually fell back and was captured and accreted. The total accretion rate onto the binary system varied periodically as the stars orbited one another, but its time-averaged value remained about the same as before binary formation. From this point onward the bubbles showed instability, and constantly changed shape while undergoing slow overall expansion. Accretion onto the system continued uninterrupted.

We halted the simulation at $57$ kyr, after a $\sim 20$ kyr period when there was no further qualitative change in the evolution (Fig.\ 1E). At this point the system was a binary with a total mass of $70.7$ $\msun$ and a time-averaged total luminosity of approximately $5\times 10^5$ $\lsun$. The two stars had masses of $41.5$ $\msun$ and $29.2$ $\msun$ and were 1590 AU apart. Neglecting the effects of the gas the semi-major axis of the orbit was $1280$ AU (eccentricity $0.25$), but because this neglects the gas, it may be an overestimate. Orbits like this are typical of young O stars, at least 40\% of which are visual binaries with separations $\sim 1000$ AU \cite{mason98a}. These are not the final system parameters, because the envelope and the disk still contained $28.3$ $\msun$ of gas and the accretion rate had not diminished. However the qualitative nature of the final system was well-established. 

We compared our result to two-dimensional simulations. The largest star that formed in any two-dimensional simulation with gray radiative transfer had a mass of 22.9 $\msun$. If the simulation included a multi-frequency treatment of the radiation, which we omitted because of its computational cost (see the SOM for details), the maximum mass of the star that formed was 42.9 $\msun$ \cite{yorke02}. In these two-dimensional simulations the initial phases of collapse, disk formation, and growth of a polar bubble were quite similar to ours, although the disk lacked non-axisymmetric structure. In both cases there was a ``flashlight effect" \cite{yorke99,yorke02} in which the disk beamed radiation preferentially into the polar direction. In two dimensions, however, as the star's mass grew, radiation halted accretion over an ever larger fraction of the solid angle around the star. This eventually stopped infall onto the disk. Some of the gas remaining in the disk continued to accrete onto the star, but at a diminishing rate, and eventually the disk density became low enough for stellar radiation to blow it away.

This never happened in our simulation. Instead, when the luminosity became large enough that our bubbles no longer delivered mass to the disk efficiently, they became asymmetric and clumpy. In some places radiation blew out sections of the bubble wall, whereas in others dense filaments of gas fell toward the stars (Fig.\ 3). The structure of dense fingers of heavy, downward-moving fluid alternating with chimneys of outgoing radiation is analogous to that of a classical Rayleigh-Taylor instability, with radiation taking the place of the light fluid. Radiation forces away from the star are stronger than gravity when averaged over $4\pi$ sr, producing velocities and net forces that have an outward direction over most of the solid angle. Much of the mass is concentrated into the dense fingers, and because radiation flows around rather than through these structures, within them the velocity and the net force have an inward direction. However, this did not remove the angular momentum of the gas, so it continued to fall onto the disk rather than directly onto the stars. The growth of clumps in the disk that form secondary stars is a natural side-effect of this process, but radiation may be just accelerating a process that is caused by gravity. At least 40\% of the accreting gas reached the disk through this Rayleigh-Taylor mechanism; gas falling onto the outer disk directly accounted for $\sim 25\%$ of the accretion, and gas reaching the disk by traveling along the bubbles' outer walls contributed the remaining $\sim 35\%$ (see SOM for details). 

Continued disk feeding is what made the three-dimensional results different from earlier two-dimensional ones. At $34.0$ and $41.7$ kyr (Fig.\ 1C, 1D), bracketing the onset of instability, the total stellar mass was 32.4 $\msun$ and 46.9 $\msun$, and the disk mass was $4.0-5.7$ $\msun$ and $4.5-9.1$ $\msun$, respectively. (The range reflects using density cutoffs of $10^{-14}-10^{-16}$ g cm$^{-3}$ to separate disk from non-disk material.) If accretion of gas onto the disk had halted, then the evolution would likely have been the same as in the two-dimensional simulations. The remnant disk would still have accreted, but its low mass means the stars would have gained less than $10$ $\msun$ from it. In our three-dimensional simulation they instead accreted $25-40$ $\msun$, and did so at a constant rather than a declining accretion rate. Accretion of stars rather than gas did not contribute significantly to this. The star with a final mass of $\sim 40$ $\msun$ gained only $1.8$ $\msun$ via collisions, while the $\sim 30$ $\msun$ star gained $1.2$ $\msun$, excluding the initial collision that created it.

The final star formation efficiency for our simulated core was at least $70\%$, and a majority of the mass accreted within in one mean-density free-fall time of the initial core, $52.5$ kyr. Because infall was continuing at a roughly constant rate and no further qualitative changes were occurring at the time we halted the simulation, it is likely that much of the remaining mass would accrete onto the star system. In reality, protostellar outflows, which we have not included in our simulation, would limit the star formation efficiency to $\sim 50\%$ \cite{matzner00,alves07}. Our result indicates that, when compared to outflows, radiation pressure does not affect star formation efficiency or timescales. The cavities generated by outflows would reduce the effects of radiation pressure even further \cite{krumholz05a}, and would modify the geometry of the radiation pressure bubbles or would prevent their formation altogether. Photon bubble instabilities, which can occur if the gas is sufficiently magnetized \cite{turner07}, might also reduce the effects of radiation pressure and modify the bubble geometry. Our simulation shows that, even if these effects are omitted, radiation pressure does not present a barrier to massive star formation.\\



\noindent
\textbf{Supporting Online Material}\\
www.sciencemag.org\\
Supporting online text\\
Figs.\ S1, S2, S3, S4, S5\\
Movie S1


\clearpage

\begin{center}
\includegraphics{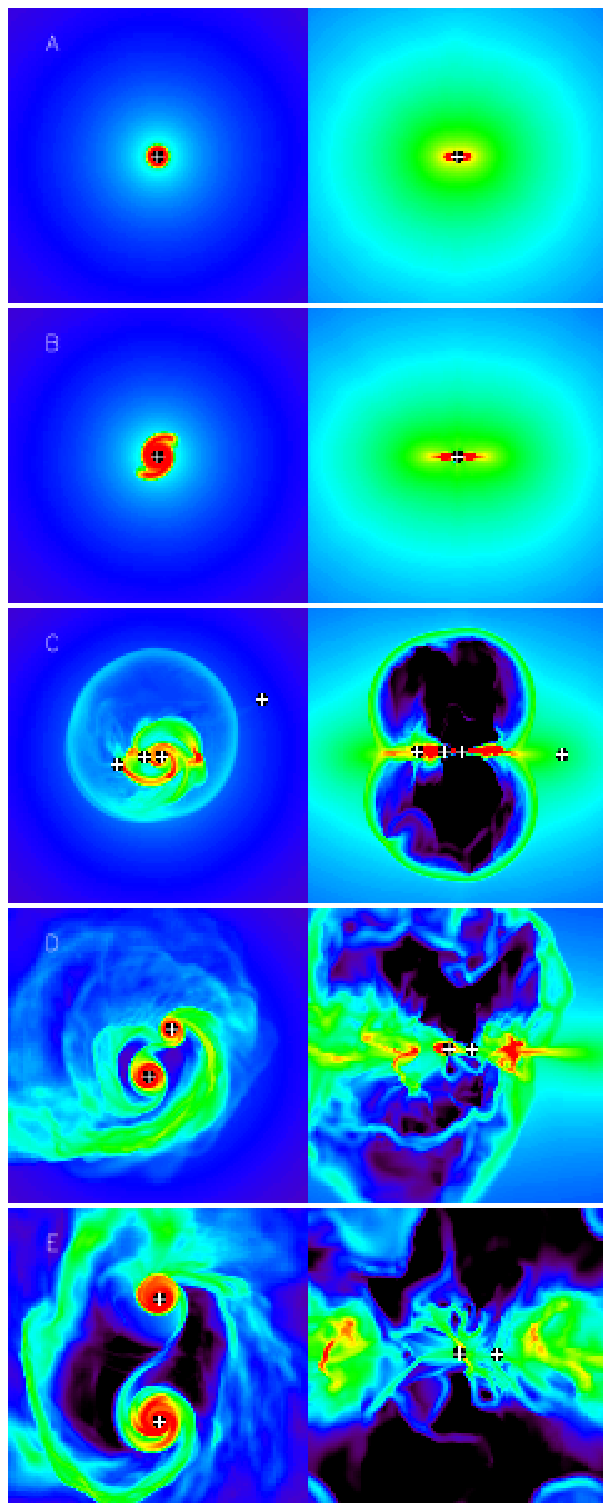}
\end{center}

\noindent {\bf Fig.\ 1.} Snapshots of the simulation at 17.5, 25.0, 34.0, 41.7, and 55.9 kyr (A--E). In each panel the left side image shows column density perpendicular to the rotation axis in a $(3000\mbox{ AU})^2$ region. The right side shows volume density in a $(3000\mbox{ AU})^2$ slice along the rotation axis.  The color scales are logarithmic, and run from $10^0-10^{2.5}$ g cm$^{-2}$ on the left and $10^{-18}-10^{-14}$ g cm$^{-3}$ on the right. Plus signs indicate the projected positions of stars. See Fig.\ S1--S3 and Movie S1 for additional images.

\clearpage

\begin{center}
\includegraphics{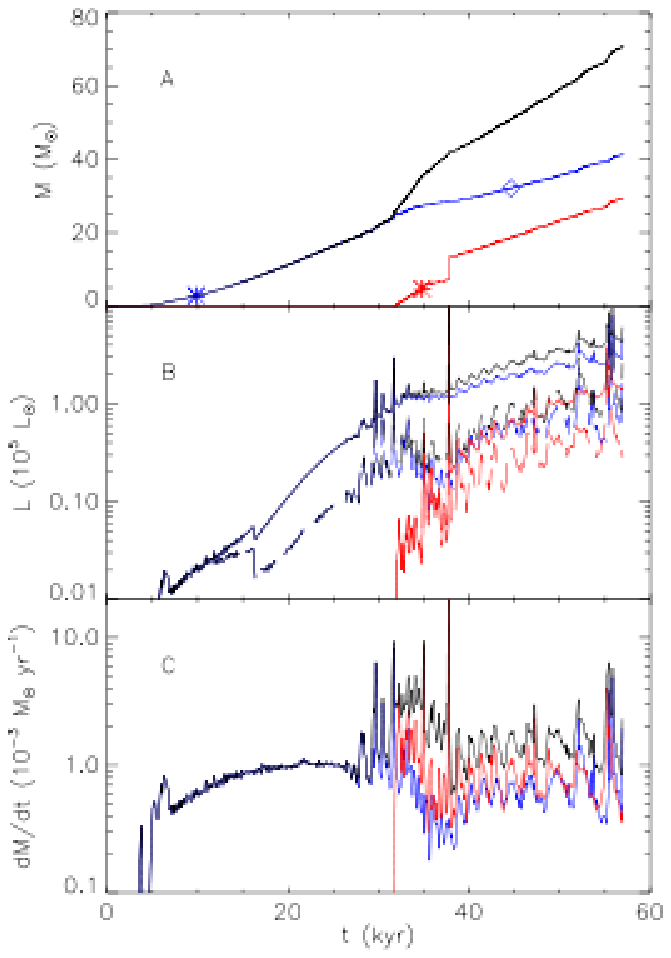}
\end{center}

\noindent {\bf Fig.\ 2.} (A) Stellar mass, (B) stellar luminosity, and (C) accretion rate as a function of time. Black lines show values summed over all stars, blue lines show values for the most massive star, and red lines show values for second most massive star. (A) Asterisks mark the onset of deuterium burning and diamonds mark hydrogen burning. (B) Solid lines show luminosities from all sources (accretion, Kelvin-Helmholtz contraction, and nuclear burning), while dashed lines show accretion luminosity only. Luminosities and accretion rates are 100 yr running averages.

\clearpage

\begin{center}
\includegraphics{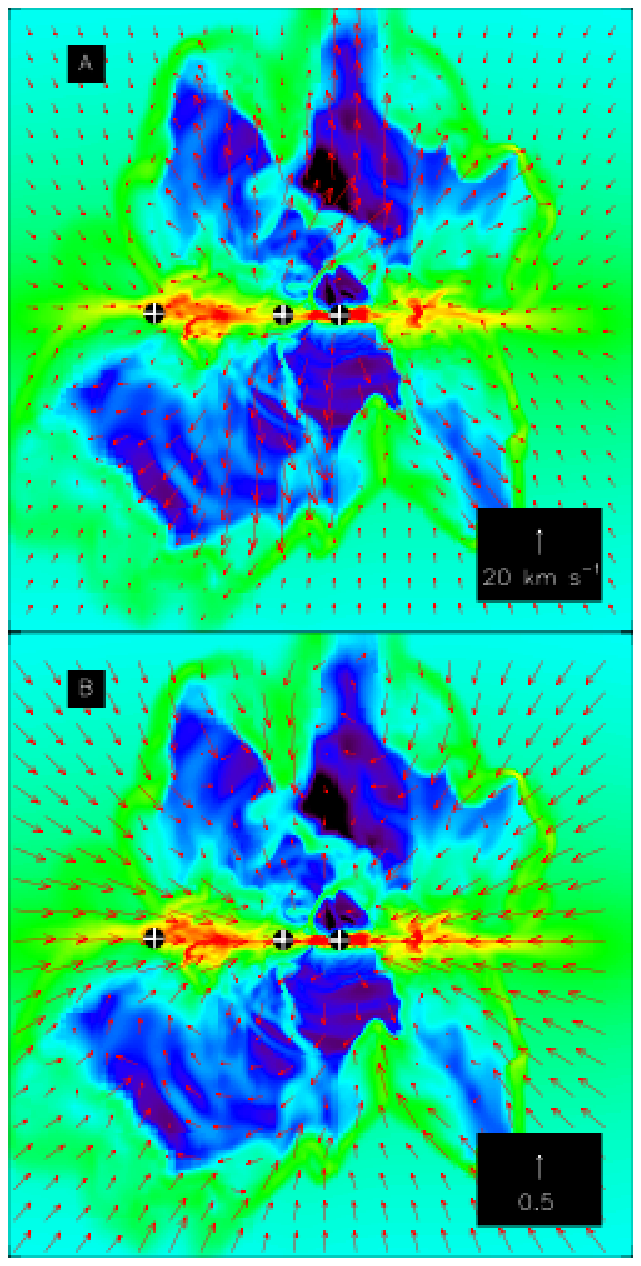}
\end{center}

\noindent {\bf Fig.\ 3.} Snapshot of a $(6000\mbox{ AU})^2$ slice along the rotation axis at 51.1 kyr. Color indicates density from $10^{-20}-10^{-14}$ g cm$^{-3}$ on a logarithmic scale. Plus signs show projected stellar positions. (A) Arrows show gas velocity. (B) Arrow directions indicate the direction of the net (radiation plus gravitational) force, while lengths are proportional to the magnitude of the net force divided by the magnitude of the gravitational force. Thus an inward arrow of length 1 represents negligible radiation force.

\setcounter{section}{0}

\clearpage

\centerline{\LARGE \bf Supporting Online Material}

\vspace{0.2in}

In this supporting text we describe our physical model, numerical method, and simulation analysis in more detail.

\section{Evolution Equations}

For clarity, in this section we write scalars in italics (e.g.\ $a$), vectors in bold (e.g.\ $\mathbf{a}$), tensor contractions over single indices as dots (e.g.\ $\mathbf{a}\cdot\mathbf{b}=a^i b^i$), and tensor products of vectors without any operator symbol (e.g.\ $(\mathbf{a}\mathbf{b})^{ij}=a^i b^j$). We write quantities evaluated in the rest frame of the computational grid without subscripts, and quantities evaluated in the frame comoving with the gas with subscript zero.

Our protostellar core is governed by the equations of gravito-radiation hydrodynamics. For the radiative part of the evolution, since the structures we form are extremely optically thick ($\tau \sim 100$ through the disk), we adopt the flux-limited diffusion approximation (see \S~\ref{radtransfer} for more details). With this approximation the state of the gas at every position is specified by a vector of quantities $(\rho, \rho\vecv, \rho e, E)$, where $\rho$ is the gas density, $\vecv$ is the gas velocity, $e$ is the gas specific energy including thermal and kinetic, but not gravitational, components, and $E$ is the radiation energy density in the rest frame of the computational grid. The computational domain also includes a number of star particles \cite{krumholz04, krumholz07a}, each of which is a point mass characterized by a mass $M$, a position $\vecx$, a momentum $\vecp$, and a luminosity $L$.

We write the evolution equations for the gas quantities in the conservative, mixed-frame form, retaining terms to order $v/c$. We use the form of the equations appropriate to the static diffusion limit, since for our problem $v/c$ is small compared to one over the optical depth of the system \cite{krumholz07b}:
\begin{eqnarray}
\dt \rho & = & -\nabla \cdot (\rho\vecv) - \sum_i \dot{M}_i W(\vecx-\vecx_i)
\label{masscons}
\\
\dt (\rho\vecv) & = & -\nabla \cdot (\rho\vecv\vecv) - \nabla P - \rho\nabla \phi - \lambda \nabla E - \sum_i \dot{\vecp}_i W(\vecx-\vecx_i)
\label{gasmom}
\\
\dt (\rho e) & = & -\nabla \cdot [(\rho e + P)\vecv] -\rho \vecv \cdot \nabla \phi - \kp \rho (4\pi B - c E) 
\nonumber
\\
& & \qquad {} + \lambda \left(2\frac{\kp}{\kr}-1\right) \vecv\cdot \nabla E - \sum_i \dot{\mathcal{E}}_i W(\vecx-\vecx_i)
\label{gasen}
\\
\dt E & = & \nabla \cdot \left(\frac{c\lambda}{\kr\rho}\nabla E\right) + \kp \rho(4\pi B - cE) - \lambda \left(2\frac{\kp}{\kr}-1\right) \vecv\cdot \nabla E 
\nonumber
\\
& & \qquad {} - \nabla\cdot\left(\frac{3-R_2}{2}\vecv E\right) + \sum_{i} L_i W(\vecx-\vecx_i).
\label{raden}
\end{eqnarray}
Equations (\ref{masscons}), (\ref{gasmom}), (\ref{gasen}), and (\ref{raden}) are the equations of mass conservation, momentum conservation, and gas and radiation energy conservation, respectively. The corresponding evolution equations for the point masses are
\begin{eqnarray}
\label{sinkmass}
\frac{d}{dt} M_i & = & \dot{M}_i \\
\label{sinkpos}
\frac{d}{dt} \vecx_i & = & \frac{\vecp_i}{M_i} \\
\frac{d}{dt} \vecp_i & = & -M_i \nabla \phi + \dot{\vecp}_i.
\label{sinkmom}
\end{eqnarray}
In these equations $\dot{M}_i$, $\dot{\vecp}_i$, and $\dot{\mathcal{E}}_i$ represent the rates at which gas mass, momentum, and mechanical (thermal plus kinetic) energy are transferred from the gas onto the $i$th point mass, $L_i$ is the luminosity of that point mass, $W(\vecx)$ is a dimensionless weighting function whose integral is unity that defines the spatial region over which transfer between the point particles and the gas occurs, and the summations in equations (\ref{masscons}) -- (\ref{raden}) and all subsequent equations run over all point masses. Equations (\ref{sinkmass}) -- (\ref{sinkmom}) describe point particles moving under the influence of gravity, while accreting mass and momentum from the gas.

The gravitational potential $\phi$ is given by the Poisson equation
\begin{equation}
\label{poisson}
\nabla^2 \phi = 4 \pi G \left[\rho + \sum_i M_i \delta(\vecx-\vecx_i)\right],
\end{equation}
and the gas pressure $P$ is given by
\begin{equation}
\label{pres}
P = \frac{\rho k_{\rm B} T_{\rm g}}{\mu} = (\gamma - 1) \rho \left(e - \frac{1}{2} v^2\right),
\end{equation}
where $T_{\rm g}$ is the gas temperature, $\mu$ is the mean particle mass, and $\gamma$ is the ratio of specific heats of the gas. We adopt $\mu=2.33 m_{\rm H}$, appropriate for a gas of molecular hydrogen and helium mixed in the standard cosmic abundance. Since over most of the computational domain the gas is too cool to excite the rotational levels of hydrogen, we approximate $\gamma=5/3$. In practice this choice makes little difference, because the gas temperature and thus the pressure are controlled almost completely by radiative rather than mechanical effects.

The quantities $\kp$ and $\kr$ represent the comoving-frame specific Planck- and Rosseland-mean opacities of the gas, which we determine from our dust grain model as described below. The quantity $B$ is the Planck function, $B=c a_{\rm R} T_{\rm g}^4/(4\pi)$. Finally, the quantities $\lambda$ and $R_2$ are the flux limiter and the Eddington factor, respectively. We adopt the Levermore \& Pomraning form for these quantities \cite{levermore81}:
\begin{eqnarray}
\label{fluxlimit}
\lambda &= & \frac{1}{R} \left(\coth R - \frac{1}{R}\right) \\
R & = & \frac{|\nabla E|}{\kr \rho E} \\
R_2 & = & \lambda + \lambda^2 R^2.
\label{eddfactor}
\end{eqnarray}

\section{Models for Dust and Protostars}

The primary source of opacity in our calculation is dust suspended in the gas. We adopt Planck- and Rosseland-mean opacities from a six-species dust model for dense interstellar environments in the Milky Way \cite{pollack94}. The opacities in the model depend on the grain temperature, since different species sublime at different temperatures. We take the grain and radiation temperatures to be equal, and the radiation temperature is given by $T_{\rm r}=(E/a_{\rm R})^{1/4}$. For convenience we fit the tabulated opacities with simple piecewise-linear analytic formulae \cite{krumholz07a}. Our fit is
\begin{eqnarray}
\kp & = &
\left\{
\begin{array}{lr}
0.3 + 7.0\,(\trk/375)
, \qquad &
\trk \leq 375 \\
7.3 + 0.7\,(\trk - 375)/200
, &
375 < \trk \leq 575 \\
3.0 + 0.1\,(\trk - 575)/100
, &
575 < \trk \leq 675 \\
2.8 + 0.3\,(\trk - 675)/285
, &
675 < \trk \leq 960 \\
3.1 - 3.0\,(\trk - 960)/140
, &
960 < \trk \leq 1100 \\
0.1, &
\trk > 1100
\end{array}
\right.
\\
\kr & = &
\left\{
\begin{array}{lr}
0.1 + 4.4\,\trk/350
, \qquad &
\trk \leq 350 \\
3.9
, &
350 < \trk \leq 600 \\
0.7
, &
600 < \trk \leq 700 \\
0.25
, &
700 < \trk \leq 950 \\
0.25 - 0.15\,(\trk - 950)/50
, &
950 < \trk \leq 1000 \\
0.1, &
\trk > 1000
\end{array}
\right..
\end{eqnarray}
For brevity we have omitted units from these equations; in them $T_{\rm r}$ is in units of K and $\kp$ and
$\kr$ are in units of cm$^2$ g$^{-1}$. At high temperatures where the
dust has sublimed, our choice to set $\kp=\kr=0.1$ cm$^2$ g$^{-1}$ is
purely a numerical convenience we use to represent a ``small''
opacity. The true opacity depends in detail on the radiation spectrum
and the physical state of the gas (molecular, atomic, or ionized); if the gas were fully ionized it would have the Thompson opacity $\kp=\kr=0.4$ cm$^2$ g$^{-1}$, but there are few places in our simulation domain that are likely to be strongly ionized. At our high accretion rates the ionized bubble around the star will be confined to scales much smaller than our grid cells \cite{keto02}. There may be significant ionization inside the radiation bubbles, where radiation-driven shocks can heat the gas to high temperatures, but the densities inside the bubbles are so low that the gas is transparent and its pressure is negligible regardless of its opacity. In the rest of the flow, gas above $1000$ K is at high density and is therefore likely to be mostly neutral. As a result it will have an opacity dominated by the lines of molecules and metal ions. These sources produce opacities that are certainly much smaller than the opacity due to dust
grains. However, sharp opacity gradients make it difficult for our
radiation iterative solver to converge, so the choice of 0.1 cm$^2$
g$^{-1}$ is a compromise between physical realism and numerical
efficiency. If anything, this choice causes us to overestimate radiative forces in parts of the flow that are above $\sim 1000$ K but neutral.

The final ingredient to our physical model is a method for determining the accretion rates and luminosities of protostars. We compute the mass accretion rate using an Eulerian sink particle algorithm
\cite{krumholz04} in which we fit the flow in the vicinity of each protostar to a Bondi accretion flow, and use the accretion rate $\dot{M}$ corresponding to the best fit. Once we have found the accretion rate, we set the momentum and energy accretion rates $\dot{\vecp}$ and $\dot{\mathcal{E}}$ so that, in the frame comoving with the accreting particle, the accretion process leaves the radial velocity, tangential momentum, and mechanical energy of the gas unchanged. We also choose the weight function $W(\vecx)$ via our Bondi flow fit, with the requirement that $W(\vecx)=0$ for $|\vecx|$ larger than four times the cell size on the finest grid. Particles can also accrete by merging with one another; this happens if one particle enters another's computationally-defined accretion radius.

Given the accretion history for each particle, we must determine its luminosity. When a sink particle first forms its mass is generally very small. Since objects smaller than $\sim 0.05$ $\msun$ do not undergo prompt collapse to stellar densities \cite{masunaga98, masunaga00}, we do not count these as stars in our analysis or allow them to radiate. Once a sink particle reaches $0.05$ $\msun$, we attach a protostellar evolution model to it \cite{mckee03, krumholz07a}. Under this model we describe the star as a polytropic sphere characterized by a radius, central temperature, and poytropic index. These evolve following the equation of energy conservation, including terms that describe the gravitational and internal chemical energy of the incoming gas, Kelvin-Helmholtz contraction of the protostar, and the energy released by deuterium and hydrogen burning. Evolution continues until the radius of the star contracts down the radius of a main sequence star of the same mass. At that point we consider the star to have ignited hydrogen and stabilized on the main sequence, and we simply take the star's luminosity to be that of a main sequence star of equal mass.

\section{Numerical Methods}
\label{numericalmethod}

We solve the evolution equations (\ref{masscons}) -- (\ref{sinkmom}) using the ORION adaptive mesh refinement (AMR) code. The code consists of four main physics modules, which operate sequentially in each update step. 

The first is the hydrodynamics module \cite{puckett92, truelove98, klein99}, which solves the Euler equations of gas dynamics (equations \ref{masscons} -- \ref{gasen}) without any of the terms involving sink particles or radiation (i.e.\ without terms involving $E$, $\dot{M}$, $\dot{\vecp}$, or $\dot{\mathcal{E}}$). This update uses a conservative Godunov scheme with an approximate Riemann solver \cite{toro97}, is second-order accurate in time and space for smooth flows, and requires very little artificial viscosity to handle shocks. The second module is the gravity solver, which solves the Poisson equation (\ref{poisson}) for the gravitational potential using a multigrid iteration scheme \cite{truelove98, klein99, fisher02}.

The third module is the radiation update, which updates the gas state based on the radiative terms (those involving $E$ and $L$) in equations (\ref{gasmom}) -- (\ref{raden}). This step uses an operator splitting approach in which we separate the terms describing radiation diffusion, heating, and cooling, which are dominant, from those describing radiation force, work, and advection, which are non-dominant \cite{krumholz07b}. We first update the state based on the dominant terms implicitly using a pseudo-transient continuation scheme \cite{shestakov08}, and then perform an explicit update for the non-dominant terms. This two-step approach is computationally cheaper than a fully implicit treatment, but remains stable and accurate as long as the force, work, and advection terms are subdominant, which they always are in our problem. We use our dust and protostellar models to compute the opacities $\kp$ and $\kr$ and protostellar luminosities $L$ in this step.

The fourth module is the star-particle update. In this step we first compute the accretion rates of mass, momentum, and energy onto each protostar and the weight function $W(\vecx)$ by fitting the density and velocity field around the star to a Bondi flow \cite{krumholz04}. We then update the properties of the gas and protostars using the terms in equations (\ref{masscons}) -- (\ref{sinkmom}) involving accretion (those involving $\dot{M}$, $\dot{\vecp}$, and $\dot{\mathcal{E}}$), and update the protostellar positions and momenta using equations (\ref{sinkpos}) -- (\ref{sinkmom}). We compute the gravitational accelerations of particles and the resulting reflex accelerations of the gas using a direct cell-by-cell and particle-by-particle calculation of the $1/r^2$ gravitational force. This ensures that our particle update conserves momentum, angular momentum, and energy to machine precision, and it is not too computationally costly because the number of particles is small. Next, we update the protostellar evolution model to determine a new set of protostellar properties for each star \cite{krumholz07a}. Finally, we create a new star in any cell whose density exceeds the Jeans density on our finest level of refinement (see \S~\ref{conditions}). Cells that meet this condition are necessarily in regions of gravitationally-bound converging flow. Rather than converting the cell entirely to a point particle, we only convert enough mass into a point particle to bring the density back below the Jeans density, although this star subsequently accretes additional mass if inflow into the cell continues. More details on our sink particle method are given in Krumholz, McKee, \& Klein \cite{krumholz04}.

All of these modules operate within the overall AMR framework \cite{berger84, berger89, bell94} in which we discretize the computational domain onto a series of levels $l=0,1,2,\ldots,L$. Here $l=0$ is the coarsest level, on which cells have a linear size $\Delta x_0$, and the cells in each subsequent level are a factor $f=2$ smaller in linear extent, so that they have sizes $\Delta x_l = \Delta x_0 / f^l$. Each level consists of a union of rectangular grids, which need not be contiguous, but which are nested, such that every grid on level $l$ is fully contained within one or more grids on level $l-1$.

Each level advances with its own time step $\Delta t^l$, but the time steps are synchronized so that $\Delta t_l = \Delta t_{l-1}/f$. The process of advancing the calculation is recursive: we first advance all grids on level 0 through a time step of size $\Delta t_0$, and then we advance all the grids on level $1$ through $f$ time steps of size $\Delta t_1 = \Delta t_0/f$. However each time we advance the level 1 grids, we must advance each level 2 grid through $f$ time steps of size $\Delta t_2 = \Delta t_1/f$, and so forth down to the finest level present. Every time we advance a level $l$ through $f$ steps, so that it reaches the same evolution time as level $l-1$, we perform a synchronization procedure between it and level $l-1$ to ensure that mass, momentum, and energy are conserved across the interface between the two levels. We set the overall time step by computing the Courant condition, including a contribution to the effective sound speed from radiation pressure \cite{krumholz07a}, on each level at the beginning of each coarse time step. We then set the time step on level 0 equal to $\Delta t_0 = \min(f^l \Delta t_l)$ and the time steps on all other levels to $\Delta t_l = \Delta t_0 / f^l$. The ensures that we obey the Courant condition on each level and that time steps on different levels are related by $\Delta t_l=\Delta t_{l-1}/f$.

\section{Initial, Boundary, and Refinement Conditions}
\label{conditions}

As described in the main text, our initial core is a 100 $\msun$ sphere of gas with a temperature of 20 K, a radius of $0.1$ pc, and a density profile $\rho\propto r^{-1.5}$. We place this core at the center of a cubical computational domain $0.4$ pc on a side. Outside the core is an ambient medium with a temperature of $2000$ K and a density equal to 1\% of the density at the edge of the core, so the core and ambient medium are in pressure balance. To prevent it from undergoing fast radiative cooling, and from inhibiting the escape of radiation from the core, we set the Planck and Rosseland opacities of the ambient medium to zero. The entire computational domain is filled with radiation with a uniform energy density $E_0=1.2\times 10^{-9}$ erg cm$^{-3}$, corresponding to a 20 K blackbody field.

We use symmetry boundary conditions for the hydrodynamics, although this has no significant effect on the calculation because no part of the core ever approaches the boundary. For the gravity module, we use Dirichlet boundary conditions, with the gravitational potential at the edge of the computational domain set equal to $-GM_{\rm c}/r$, where $M_{\rm c}=100$ $\msun$ is the core mass, and $r$ is the distance of a given point on the boundary of the computational domain from the core center. The radiation module uses a Marshak boundary condition, under which the flux out of the computational domain at the face of each cell at the boundary is set equal to $c(E-E_0)/4$, where $E$ is the energy density in that cell. This effectively bathes the computational domain in a 20 K background radiation field, but allows excess radiation generated within it to escape freely.

Our level 0 base grid has $128^3$ cells, and we refine any cell with a density greater than half the initial edge density of the core to level 1. Thus the initial core is resolved by at least 64 cells per core radius. Thereafter we refine based on three conditions. The first, and by far the most restrictive, is that we refine any computational cell on level $l$ in which 
\begin{equation}
\frac{\Delta x_l |\nabla E|}{E} > 0.15,
\end{equation}
i.e.\ anywhere the local gradient in the radiation energy density exceeds 15\% per cell. We find that artificial radiation pressure forces develop whenever steep gradients in the radiation energy density coincide with boundaries between AMR levels, and this condition prevents such forces from appearing by guaranteeing that regions of sharp gradients in $E$ never coincide with level boundaries. The second condition is that we refine any cell whose distance from the nearest sink particle is less than $16\Delta x_l$, so that the regions around stars are always well-resolved. Finally, we refine any cell whose density is high enough to violate the Jeans condition,
\begin{equation}
\rho > \rho_{\rm J} = J^2 \frac{\pi c_{\rm s}^2}{G \Delta x_l^2},
\end{equation}
where we use a Jeans number $J=1/8$, and $c_{\rm s}$ is the isothermal sound speed of the gas.

We allow refinement to continue up to a maximum level $L=6$, at which the cell size is $\Delta x_L = 10$ AU, giving an effective resolution of $8192^3$. Cells whose densities exceed $\rho_{\rm J}$ at the highest level of resolution form star particles \cite{krumholz04}. We take no action when one of the other refinement conditions is violated on the highest level. For the radiation gradient condition this causes no harm, because the condition's primary purpose is to ensure that there are no level boundaries created by other conditions that coincide with sharp gradients in $E$. For the condition that we always refine the volume around sink particles, the consequence of our failure to do this indefinitely is that very close to a sink particle we start to suffer from artificial angular momentum loss, and that we cannot resolve sink particle orbits that are below a certain size. We discuss these issues below.

With these choices of resolution and refinement condition, the simulation we present in this paper required approximately 250,000 CPU-hours, running in parallel on $128$ or $256$ processors.

\section{Resolution Issues}

Our minimum cell size of 10 AU has important implications for which physical processes we can and cannot resolve. As we discuss in \S~\ref{instrefine}, this cell size is more than adequate for the radiation dominated-bubbles and radiation-Rayleigh-Taylor fingers. However, there are three important physical processes that our simulations cannot address due to limited resolution, which we discuss in \S~\ref{dustdestruct} -- \ref{closebinary}.

\subsection{The Dust Destruction Front}
\label{dustdestruct}

One problem we cannot address is the structure of the dust destruction front. Close to the star the temperature of a dust grain in equilibrium with the radiation field exceeds the sublimation temperature for even the most refractory species, so no dust is present. As a result the opacity of the gas to non-ionizing radiation is small, and stellar visible and ultraviolet light can free-stream outward. As one moves away from the star the equilibrium dust temperature drops below the sublimation temperature, and dust re-appears. Since the dust opacity to visible and UV light is very large, all the stellar radiation is absorbed in a thin layer at this dust destruction front, where it is down-converted to infrared and re-radiated to diffuse outward through the core. We refer to the transition layer in which direct stellar radiation is reprocessed and dust is destroyed as the dust destruction front.

The radius of the front increases with the stellar luminosity as roughly $r\propto L^{1/2}$, but even at our peak luminosity of a few $10^5$ $\lsun$ it is at most $\sim 50$ AU \cite{wolfire87}, with a far smaller thickness. In our simulation we do find a $\sim 50$ AU region at the center of the accretion disk in which the temperature is well above $1000$ K, the temperature at which our dust model opacity falls to its floor, but at our resolution of 10 AU this region is at best marginally resolved. The boundary of the high temperature region, where UV and visible radiation are reprocessed, is not resolved at all. As a result, we cannot study the structure of the front, nor can we address the ``UV" radiation pressure problem \cite{mckee07b}: at the front the large UV / visible opacity makes the radiation pressure force very large. Since the front is of very small thickness, the force is applied over a very short length and time, so it can be approximated as delivering an impulsive kick to the gas. If the accretion flow has insufficient inward momentum, the kick may reverse it. However, analytic calculations of this effect show that if the accretion rate is $\sim 10^{-4}$ $\msun$ yr$^{-1}$ or higher, as it is for all of our simulation (Fig.\ 2 of main text), then even if it is spherically symmetric the inflow always contains enough momentum to fall through the dust destruction front and onto the star \cite{wolfire87}. The non-sphericity of our inflow, which arrives primarily through a disk, will further relax this requirement, since the inflowing gas will intercept only a fraction of the stellar radiation.

Finally, it is worth noting that, even with better resolution, a realistic simulation of the dust destruction front would require a considerably more sophisticated approach to radiative transfer than the gray flux-limited diffusion approximation we use. Simulating the dust destruction front requires the ability to treat a radiation field that is non-blackbody and highly beamed, neither of which is possible with a gray flux-limited diffusion approximation. (We discuss these issues in more detail in \S~\ref{radtransfer}.) No simulation to date has studied the dust destruction front even in 2D. The closest are those of Yorke \& Sonnhalter \cite{yorke02}, which include a multi-frequency treatment of the radiation field, one of the necessary components. However, they too use a diffusion method and thus cannot handle beamed radiation fields, and their simulations also have significantly lower resolution than ours, so the entire dust destruction front is confined to a single cell of their simulation.

\subsection{Angular Momentum Transport and Fragmentation in the Disk}

Our maximum resolution of 10 AU also sets the size scale outside which our disks are dominated by physical angular momentum transport due to gravitational instability, and inside which they are dominated by numerical angular momentum transport induced by finite resolution. Numerical angular momentum transport leads to artificially fast accretion in regions of the disk where it dominates, which in turn produces an unphysically smooth and low surface density disk that is less prone to fragmentation than it should be.

We have measured the effective numerical viscosity in our code experimentally \cite{krumholz04}. Parameterizing this in terms of the standard viscosity parameter $\alpha$, our effective grid viscosity for a disk in Keplerian orbit is \cite{krumholz07a}
\begin{equation}
\alpha \approx 3.5 \Delta x_{L,1}^{-2.85} M_0 T_{2}^{-1} r_2^{-3.85}
\end{equation}
where $\Delta x_{L,1}=1$ is the size of a cell on our finest level $L$ in units of 10 AU, $M_0$ is the mass of the central star in units of $1$ $\msun$, $T_2$ is the gas temperature in the disk in units of $100$ K, and $r_2$ is the distance from the central star in units of $100$ AU. To determine when this is important, we must compare it to the dominant source of physical angular momentum transport: gravitational instabilities that produce spiral arms. These can produce an effective $\alpha\sim 1$ for a rapidly-accreting disk \cite{krumholz07a, kratter08a}. Since $M_0/T_2\sim 1$ in our disks for most of the simulation, we expect numerical angular momentum transport to become negligible compared to physical transport at radii larger than about $100$ AU. 

Because of this effect, we likely overestimate the accretion rate and underestimate the amount of fragmentation and spiral structure at early times, when the disk radius is small. As the run continues, however, the solid-body rotation curve we impose as an initial condition causes the circularization radius of the infalling material to grow. As the time-sequence of the evolution shows, by the time radiation effects become significant, the disk is several hundred AU in radius, and has a well-developed spiral arm-pattern, indicating that physical angular momentum transport has become dominant and numerical viscosity is negligible. At this point, even though we are still overestimating the accretion rate in the inner $\sim 100$ AU of the disk, this does not raise the total accretion rate. Simulations exploring the effects of numerical viscosity and artificial accretion in disks show that even completely evacuating the inner, numerical viscosity-dominated region does not cause any significant accretion from the outer parts if they are stable and rotationally-supported \cite{krumholz04}. Physically this is because the rate-limiting step for accretion is the delivery of mass from the outer parts of the disk to their inner parts. Speeding up accretion in the inner part of the disk does not remove this bottleneck.

It is also worth noting that none of this discussion of numerical viscosity applies to the star particles that we create in our simulation. Numerical viscosity in our code arises for the same reason it does in any numerical method, either grid- or particle-based: fluid properties are always averaged over some finite size scale, and when the averaging scale becomes comparable to the orbital radius of a given fluid element, either artificial angular momentum transport or artificial angular momentum loss occurs. However, our star particles are in fact point masses and not fluid elements of finite size, so they do not suffer from the averaging problem. As discussed in \S~\ref{numericalmethod} of the Supplementary Text, we use an operator-split approach to update the gas and the particles, and the particle update conserves angular momentum to machine precision. Thus our sink particles do not suffer from numerical viscosity. They do interact with the gas gravitationally, however, so even though their motions are dissipationless, they can still be dragged inward by dynamical friction with the gas. This is the dominant accretion mechanism for disk-born stars, which all form outside the $\sim 100$ AU zone where numerical viscosity affects the gas. Until formation of the massive binary companion, all the stars in th disk are too small to open gaps in the disk. The disk itself is accreting rapidly due to gravitational instability, so the stars undergo rapid type I migration inward.

\subsection{Close Binaries}
\label{closebinary}

The final area in which our resolution compromises our ability to model the physics is in our treatment of tight binaries. O star binaries are generally detected either visually \cite{mason98a}, spectroscopically \cite{sana08a}, or by eclipses \cite{harries03, hilditch05}. Visual binaries are easiest to find if they have separations of at least $\sim 100$ AU, while spectroscopic and eclipse techniques are most sensitive to binaries closer than $\sim 1$ AU. Companions to O stars in the intermediate regime are extremely difficult to detect, and we lack a good census for them. Surveys indicate that $\sim 40\%$ of O stars have companions in the visual binary regime \cite{mason98a}, with $\sim 1000$ AU as a typical separation, while at least 60\% have companions in the spectroscopic / eclipse regime \cite{sana08a}. Our sink particle algorithm merges stars whose 40 AU-sized accretion zones overlap, so while we resolve visual companions easily, closer spectroscopic or eclipsing binaries are unresolved in our simulation. As a result, we cannot model the formation of tight binaries or make any statements about how common we expect such systems to be.

Both the first star to form in our simulation and its disk-borne wide companion undergo a number of mergers, although these contribute only a small fraction of their total mass. It seems likely that at least some of the mergers should instead result in the formation of close binaries below our resolution limit. The net effect of this would be to reduce the relative strength of radiation pressure. Far away from such a tight binary the gravitational force is the same as it would be for a single star of the same mass, but the strong non-linear dependence of luminosity on mass means that two 20 $\msun$ stars in a tight binary produce less light than a single 40 $\msun$ star. Consequently, our limited resolution means that, if anything, we overestimate the radiation pressure force. Since we find that radiation cannot stop accretion even in this case, a higher-resolution treatment of tight binaries would only strengthen our result.

\subsection{Resolution of Bubbles and Rayleigh-Taylor Features}
\label{instrefine}

In contrast to the issues of the dust destruction front, inner disk, and close binaries, our resolution is more than adequate to simulate the radiation-dominated bubbles and the appearance of Rayleigh-Taylor instabilities within them. We can quantify this two ways. Most simply, we note that, at the onset of instability in the bubbles above and below the accretion disk, they are $\sim 1500$ AU in diameter (Fig.\ 1C--1D of main text). Since at the onset of instability the entire bubble wall is refined to the maximum level, this means that the bubble circumference is resolved by $\sim 500$ cells. Numerical studies of classical (as opposed to radiative) Rayleigh-Taylor instability find that perturbation growth rates converge once the perturbations are resolved by roughly 8 cells per wavelength \cite{dimonte04a}. It not entirely clear what perturbations play a dominant role in initiating Rayleigh-Taylor instability in our problem, but examination of Movie S1 suggests they have sizes that are comparable to the bubble radius, and thus should be extremely well-resolved by our $\sim 500$ cells. If there are smaller modes that we are missing due to inadequate numerical resolution, then our Rayleigh-Taylor instability sets in too slowly in the simulation, and our conclusion that instability allows continuing accretion would be strengthened.

One could also worry about our refinement criteria, and in particular whether mesh refinement could be initiating the Rayleigh-Taylor instabilities we see. We note first that, even if this were true, it would not necessarily make our results unphysical, since in any real collapsing core there will be inhomogeneities present that are much larger than those induced by our mesh refinement. However, we can also check this possibility simply by overlaying the mesh refinement on top of the density field to look for correlations between instability features and level boundaries. Fig.\ S4 shows such an overlay; there is clearly no correlation between features of the instability and boundaries between AMR levels. Regions of inflow and Rayleigh-Taylor fingers do not correlate with resolution. While this is not definitive proof that our instability is not numerical in origin, it does provide strong circumstantial evidence against the hypothesis.

\section{Limitations of the Radiative Transfer Methodology}
\label{radtransfer}

Our simulation uses a gray flux-limited diffusion approximation for radiation. This is necessary to render the calculation computationally feasible, and is a reasonable approximation for a first three-dimensional calculation, but it is important to consider the extent to which this might influence our result. We consider the grayness and diffusion approximations separately.

The assumption behind a gray treatment of the radiation field is that the radiation always has a thermal spectrum, and to the extent that it is not fully thermalized our approximation will be in error. The error is likely to be largest close to the star, where the radiation has not yet passed through much dust, but deviations from a blackbody spectrum will be present everywhere. Dust opacity is highest at short wavelengths, so short-wavelength radiation is both the most effective at delivering momentum to the gas and the most easily collimated by it. Conversely, the core is optically thin to radiation whose wavelength is sufficiently long, and this radiation will escape, increasing the cooling rate but also hardening the spectrum of the remaining, trapped radiation. As a result, gray treatments of the radiation have two effects on the dynamics, which act in opposite directions. First, by assuming instantaneous thermalization of the radiation field, they artificially soften the radiation spectrum and underestimate the radiation force in regions where the true spectrum is harder than thermal \cite{preibisch95}. This is worst close to the star, and it artificially favors accretion. On the other hand, the assumption of instantaneous thermalization reduces the energy loss rate by neglecting the escape of long wavelength radiation, and also tends to isotropize the radiation field artificially, because it neglects the extra collimation provided by high opacities at short wavelengths. Both effects tend to make accretion more difficult, since the former raises the amount of radiation trapped in the core and the latter makes it harder for the gas to form self-shielding structures that exclude the radiation field like the Rayleigh-Taylor fingers we find in our simulation.

The best way to assess which of these effects is dominant is to compare the results of previous {\it two-dimensional} simulations with gray and with multi-group radiative transfer \cite{yorke02}. (In a multi-group approach, one integrates the radiation-hydrodynamic equations over some number of intervals in frequency space, and tracks each frequency group independently.) These show that a multi-group treatment of the radiation field in 2D increases the maximum stellar mass that is formed from 22.9 $\msun$ to 42.9 $\msun$, so in 2D artificial trapping and isotropization of the radiation field under the gray assumption is much more important than underestimation of the radiation force by artificial spectral softening. While only a 3D multi-group simulation can fully answer the question of whether this continues to be true in three dimensions, it seems likely. Trapping and softening are both spectral effects and not geometric ones, so it is unclear why they would be sensitive to dimensionality. For the isotropization effect, if anything the 3D simulation is more vulnerable than the 2D ones. In 2D, the only collimation that occurs is by the disk and the walls of the polar cavities, which are relatively large-scale features that are reasonably optically thick even to thermalized radiation. In contrast, in 3D we also have collimation by Rayleigh-Taylor fingers, which are much smaller in scale, less optically thick, and therefore more vulnerable to artificial leakage of radiation due to an underestimate of their opacity by the gray approximation. Thus the amount of extra collimation provided by a frequency-dependent treatment should be at least as great in 3D as in 2D, and multi-group radiative transfer would again enhance accretion.

The impact of the diffusion approximation is more difficult to assess, because there have been no 2D simulations of massive star formation using radiative transfer methods other than diffusion. Since the core is extremely optically thick even at infrared wavelengths, the radiation field should be close to the diffusion limit everywhere except near the core surface and inside the dust destruction front. Near the core surface beaming is unlikely to be important, since radiation forces are small there and we care only about the total energy loss rate, which the flux-limited diffusion approximation computes reasonably well. Within the dust destruction front beaming would be a more significant effect. The weight function $W(\mathbf{x}-\mathbf{x}_i)$ we use to add radiation from our stars to the grid is spherically-symmetric, so it adds radiation in and outside of the disk in an unbiased manner. However, that radiation is then free to diffuse, and little of its diffuses through the disk because of the disk's high optical depth. In reality, within the dust destruction front the radiation should be highly beamed, and this probably results in more radiation striking the inner edge of the disk than our diffusion method would compute. However, as noted in \S~\ref{dustdestruct} of the Supplementary Text, this ``UV" radiation pressure problem is not significant if the accretion rate is even close to as high as we are finding. Moreover, since we lack the resolution to study the dust destruction front in any event, our omission of radiation beaming there does not worsen the situation.

\section{What Fraction of Mass is Processed by Radiative Rayleigh-Taylor Instability?}

Once radiation bubbles and the associated Rayleigh-Taylor instability forms, gas can reach the accretion disk and thence the stars in three ways. First, gas can fall into the equatorial plane and then strike the outer rim of the disk before it encounters the radiative shock. Second, it can strike the shock, then slide along the outer wall of the radiative bubble until it reaches the disk. Third, it can strike the shock, penetrate into the bubble interior as part of a Rayleigh-Taylor finger, and then reach the disk. The distinction between these channels is somewhat fuzzy, but since only the third channel represents true Rayleigh-Taylor instability, assessing the importance of Rayleigh-Taylor instability requires that we roughly estimate their relative contributions.

To do so we compute the mass flux across three surfaces. First, we draw a spherical shell of radius $r_1$ centered on the stars' center of mass that surrounds the radiative shock as closely as possible. We estimate the angle $\alpha_e$ around the equatorial plane over which incoming gas can strike the disk directly rather than passing through the radiative shock. We refer to the region within angle $\alpha_e$ of the equatorial plane as the equatorial region, and the region outside this is the polar region (Fig.\ S5). We compute the inward mass flux through the equatorial region via
\begin{equation}
\dot{M}_{\rm eq} = -2\pi r_1^2 \sin \alpha_e \frac{\sum \rho v_r \Delta V}{\sum \Delta V},
\end{equation}
where $v_r$ is the radial velocity, $\Delta V$ is the cell volume, and the sums run over all cells within an angle $\alpha<\alpha_e$ of the equatorial plane and whose radii are within $2.5\%$ of $r_1$, and which are not covered by cells on a finer AMR level. (This last restriction is to avoid double-counting the same volume that is represented on multiple AMR levels.) We compute the mass flux through the polar region $\dot{M}_{\rm po}$ in an analogous fashion, by summing over cells with $\alpha > \alpha_e$. The ratio $\dot{M}_{\rm eq} / (\dot{M}_{\rm eq}+\dot{M}_{\rm po})$ gives the fractional contribution of gas that strikes the disk directly, without encountering the shock.

Second, we draw a shell of radius $r_2$, again centered at the stars' center of mass, chosen to be as large as possible subject to the constraint that the shell be entirely contained inside the bubble interior, and exclude the points where the bubble walls connect to the disk. Within the interior of this sphere we construct a pair of conical surfaces, which make angles of $45^{\circ}$ relative to the equatorial plane and meet at the sphere's center (Fig.\ S5). The cones represent a rough estimate of the ``disk surface". We compute the mass flux through them via
\begin{equation}
\dot{M}_{\rm ds} = -\sqrt{2}\pi r_2^2 \frac{\sum \rho v_n \Delta V}{\sum \Delta V},
\end{equation}
where $v_n$ is the component of the velocity normal to the cone surface (which is simply the negative $\theta$ component of the velocity for the upper cone, and the positive $\theta$ component for the lower cone), and the sum runs over all cells at positions $(x,y,z)$ whose angles $\theta=\tan^{-1} (z/\sqrt{x^2+y^2})$ relative to the polar axis are within 20\% of $45^{\circ}$. We also exclude cells at radii $r=\sqrt{x^2+y^2+z^2}<100$ AU from the sum, since our resolution on such small scales is poor. 
The ratio $\dot{M}_{\rm ds}/\dot{M}_{\rm po}$ tells us what fraction of the mass that strikes the radiative shock reaches the disk surface via Rayleigh-Taylor fingers inside the bubble, rather than through the bubble walls. (Strictly speaking this assumes that the mass flux is constant, since we are computing $\dot{M}_{\rm ds}$ at a smaller radius than $\dot{M}_{\rm po}$ and $\dot{M}_{\rm eq}$, but since the accretion rate on large scales varies slowly compared to the time required to travel from $r_1$ to $r_2$, this effect is small.)

There are two important points to note about this definition of the flux through the disk surface. The first is that the $45^{\circ}$ opening angle we assume for the disk surface is much larger than the opening angle $\alpha_e\approx 10^{\circ}$ we use to distinguish the equatorial from the polar region when discussing the outer disk. The reason is clear from Fig.\ S5: the inner disk is strongly flared due to radiative effects, so its scale height is larger than that of the outer disk beyond the bubble walls. A cone with an opening angle of $\sim 10^{\circ}$ is above the disk surface outside the bubble, but not inside it. That said, our result is not very sensitive to the choice of $45^{\circ}$; varying this by $\sim 10^{\circ}$ has little effect on the qualitative conclusion. The second point is that $\dot{M}_{\rm ds}$ probably represents only a lower limit on the mass accreted through the disk surface in the bubble interior. That is because we must choose $r_2$ so that it is smaller than the smallest radius at which the bubble outer wall hits the disk surface. However, the bubble is not cylindrically symmetric, so there are regions of disk surface which are well inside the radiative bubble but which are outside $r_2$. Our estimate for $\dot{M}_{\rm ds}$ does not include accretion onto these regions of the disk surface through Rayleigh-Taylor instability, so it is only a lower limit.

At 51.1 kyr, the time illustrated in Fig.\ S5, this procedure yields mass fluxes $\dot{M}_{\rm po}=1.5\times 10^{-3}$ $\msun$ yr$^{-1}$, $\dot{M}_{\rm eq} = 5.1\times 10^{-4}$ $\msun$ yr$^{-1}$, and $\dot{M}_{\rm ds} = 7.9\times 10^{-4}$ $\msun$ yr$^{-1}$. Thus, we find that 75\% of the incoming mass encounters the radiative shock rather than accreting directly onto the disk. Of this, at least 52\% reaches the disk via Rayleigh-Taylor fingers that pass through the disk surface, rather than by traveling along the shock. This calculation uses $r_1=3900$ AU, $r_2=1100$ AU, and $\alpha_e=8^{\circ}$ (Fig.\ S5). Varying these choices within reasonable limits causes these numbers to vary by only $\sim 10\%$. We therefore conclude, that, at the time shown in Fig.\ S5, accretion via Rayleigh-Taylor fingers is the largest single mode, accounting for 40\% of the total accretion. Accretion of gas that slides along the bubble walls is second, providing 35\% of mass flux, and accretion of gas the reaches the disk without encountering the shock is third at 25\%. As noted above, since the disk surface flux we calculate is a lower limit, the true Rayleigh-Taylor contribution is probably somewhat larger than 40\%, and the bubble wall contribution somewhat lower than 35\%.

At first this large flux via Rayleigh-Taylor fingers might seem surprising, given the relatively low density inside most of the disk surface region shown in Fig.\ S5. However, the appearance is somewhat deceptive. First, the velocity of gas passing through the disk surface is typically significantly larger than the velocity of gas passing through the polar and equatorial regions further out, and this partly compensates for the lower density. Second, most of the accretion through the disk surface arrives via thin filaments, and because these are wrapped around the polar axis by rotation, only small segments of a few filaments are visible in any given slice. For example, in the upper panel of Fig.\ S5 one can see a filament of dense gas passing through the upper left cross-hatched region. However, the volume rendering in Fig.\ S3 makes it clear that there are many such filaments, each contributing to the mass flux onto the disk surface. Any single slice misses most of these.

Performing the same exercise at other times after the onset of Rayleigh-Taylor instability (between the times shown in Fig.\ 1D--1E) yields results that are either similar or show an even larger fraction of the mass reaching the disk via Rayleigh-Taylor instability. In contrast, the results are quite different before the noticeable onset of Rayleigh-Taylor instability. For example at 34.0 kyr, the time shown in Fig.\ 1C, we find $\dot{M}_{\rm po} = 1.6\times 10^{-3}$ $\msun$ yr$^{-1}$, $\dot{M}_{\rm eq}=4.5\times 10^{-4}$ $\msun$ yr$^{-1}$, and $\dot{M}_{\rm ds}=1.9\times 10^{-4}$ $\msun$ yr$^{-1}$. Thus only $\sim 10\%$ of the mass reaching the radiative shock accretes through the disk surface inside the bubble, while the remaining $\sim 90\%$ reaches it through the bubble walls. This changes qualitatively between 34.0 and 41.7 kyr, when visible Rayleigh-Taylor fingers appear.

\clearpage

\begin{center}
\includegraphics{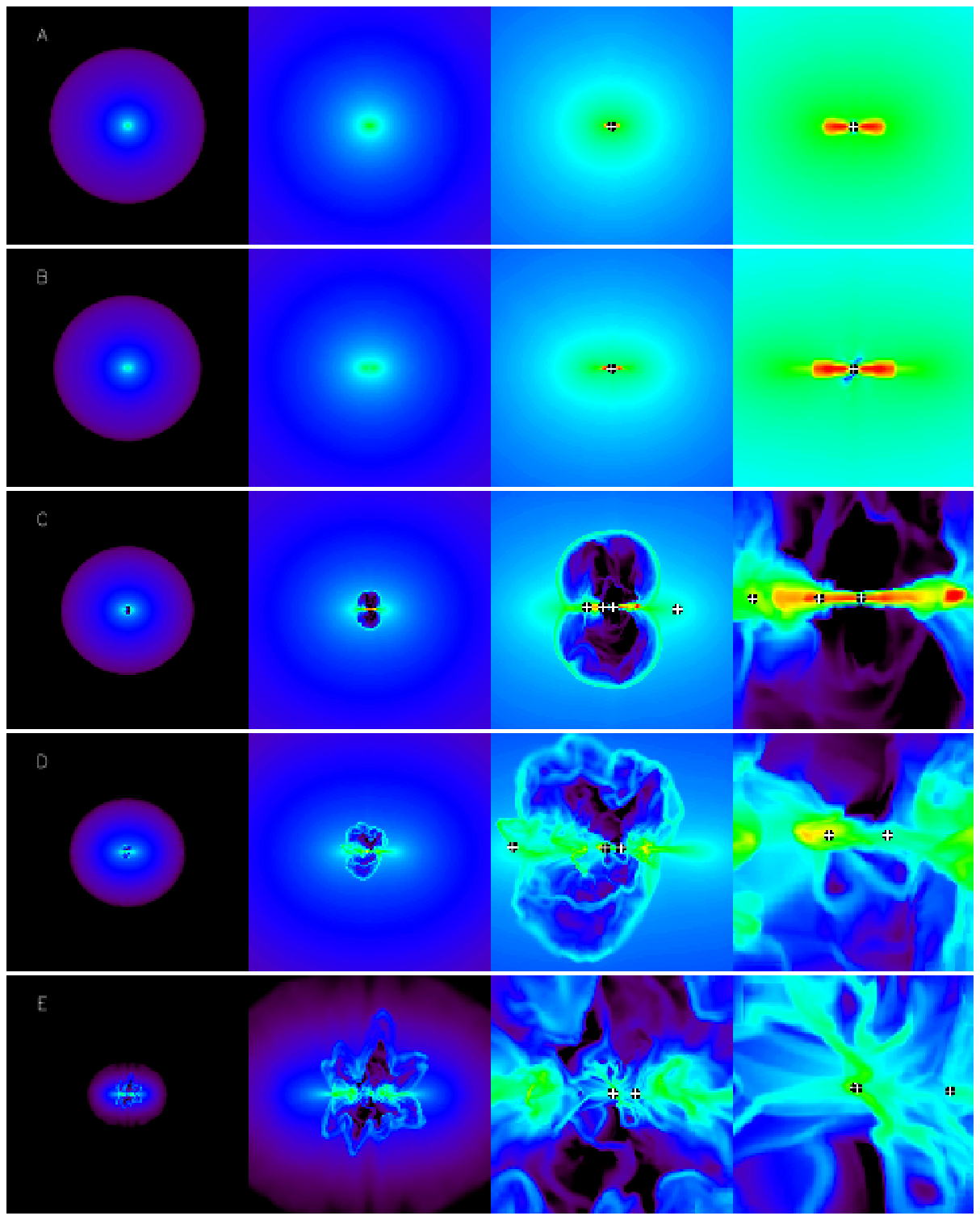}
\end{center}

\noindent {\bf Supp.\ Fig.\ 1.} Five snapshots of the simulation domain at the same times as those shown in Fig.\ 1 of the main text: (A) $17.5$ kyr, (B) $25.0$ kyr, (C) $34.0$ kyr, (D) $41.7$ kyr, (E) $55.9$ kyr. In each panel the four frames all show slices of density through the simulation domain in a plane along the rotation axis. The leftmost frame shows a $(0.3\mbox{ pc})^2$ region, and each step to the right reduces the size of the region shown by a factor of 4 in linear dimension, so that the rightmost box shows a region $(966\mbox{ AU})^2$ in size. The color scale is logarithmic, running from $10^{-19} - 10^{-12}$ g cm$^{-3}$. Plus signs mark projected star positions. For clarity we only show stars in the rightmost two frames.

\clearpage

\begin{center}
\includegraphics{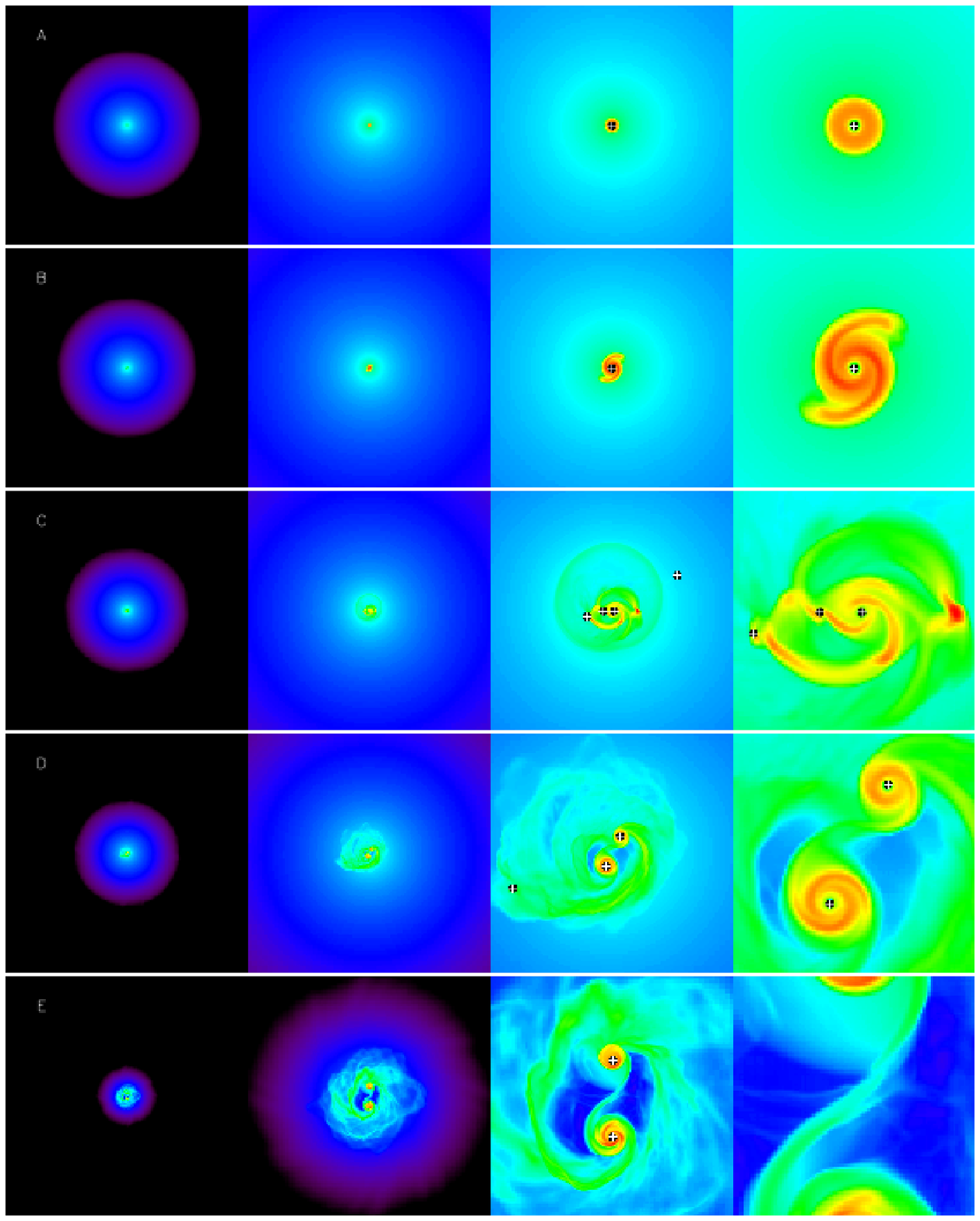}
\end{center}

\noindent {\bf Supp.\ Fig.\ 2.} Snapshots of the simulation domain at the same times as Fig.\ 1 and Fig.\ S1. In each panel the four frames all show the column density projected onto a plane orthogonal to the rotation axis. The leftmost frame shows a $(0.3\mbox{ pc})^2$ region, and each step to the right reduces the size of the region shown by a factor of 4 in linear dimension, so that the rightmost box shows a region $(966\mbox{ AU})^2$ in size. The color scale is logarithmic, running from $10^{-1} - 10^{3}$ g cm$^{-2}$. Plus signs mark projected star positions. For clarity we only show stars in the rightmost two frames.

\clearpage

\begin{center}
\includegraphics[scale=0.6]{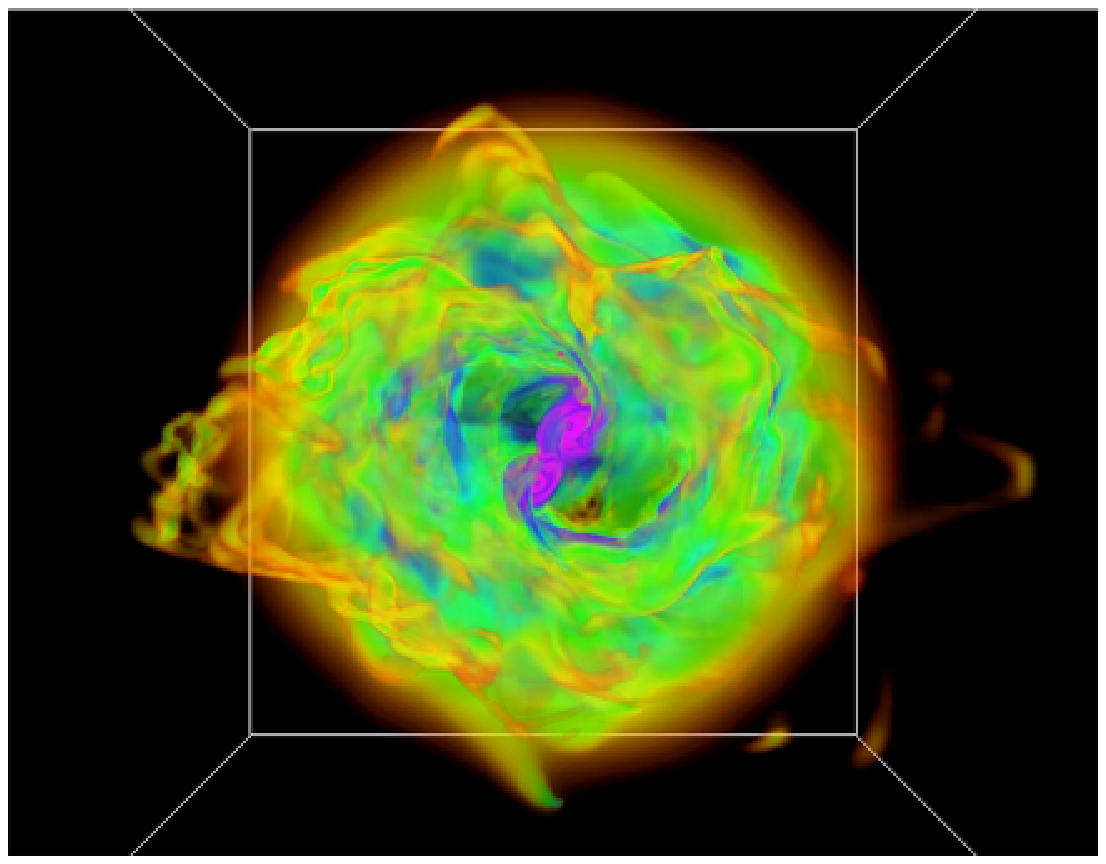}\includegraphics[scale=0.6]{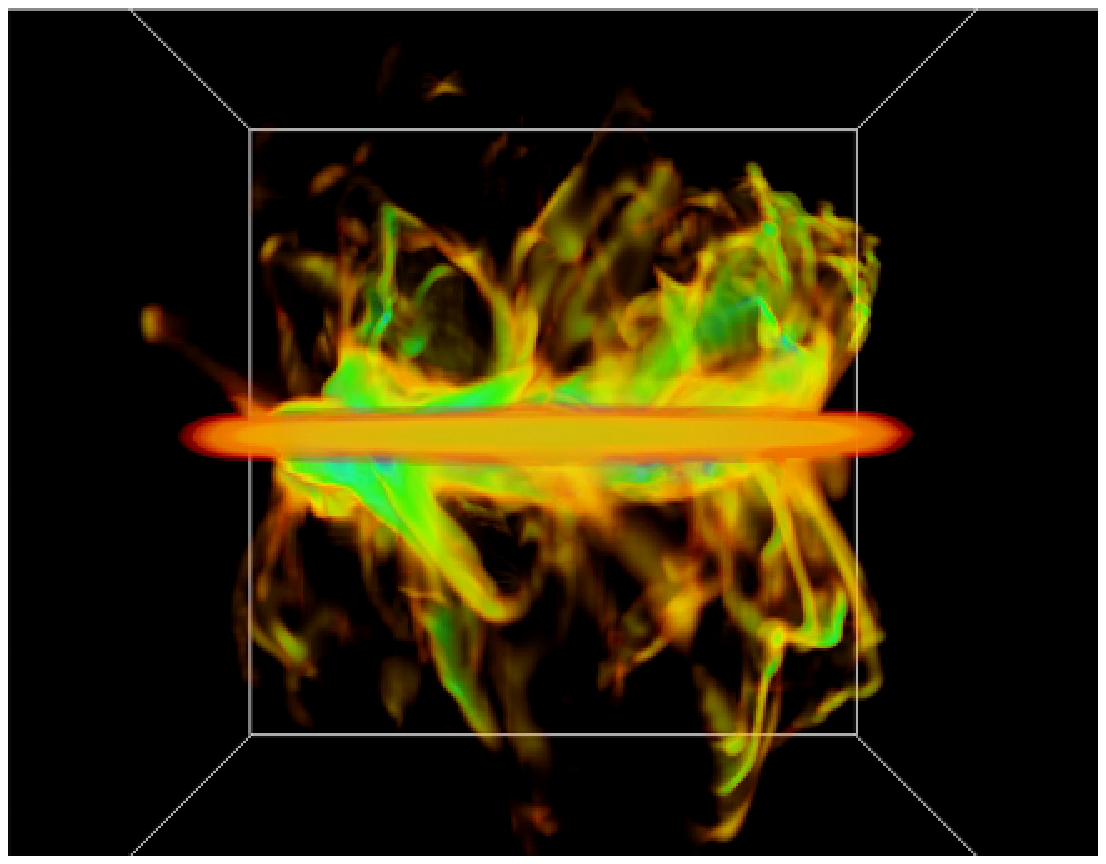}
\end{center}

\noindent {\bf Supp.\ Fig.\ 3.} Volume renderings of the density field in a $(4000\mbox{ AU})^3$ region of the simulation at 55.0 kyr of evolution. The color scale is logarithmic and runs from $10^{-16.5}-10^{-14}$ g cm$^{-3}$. The left panel shows a polar view, and the right panel shows an equatorial view. The Rayleigh-Taylor fingers feeding the equatorial disk are clearly visible.

\clearpage

\begin{center}
\includegraphics{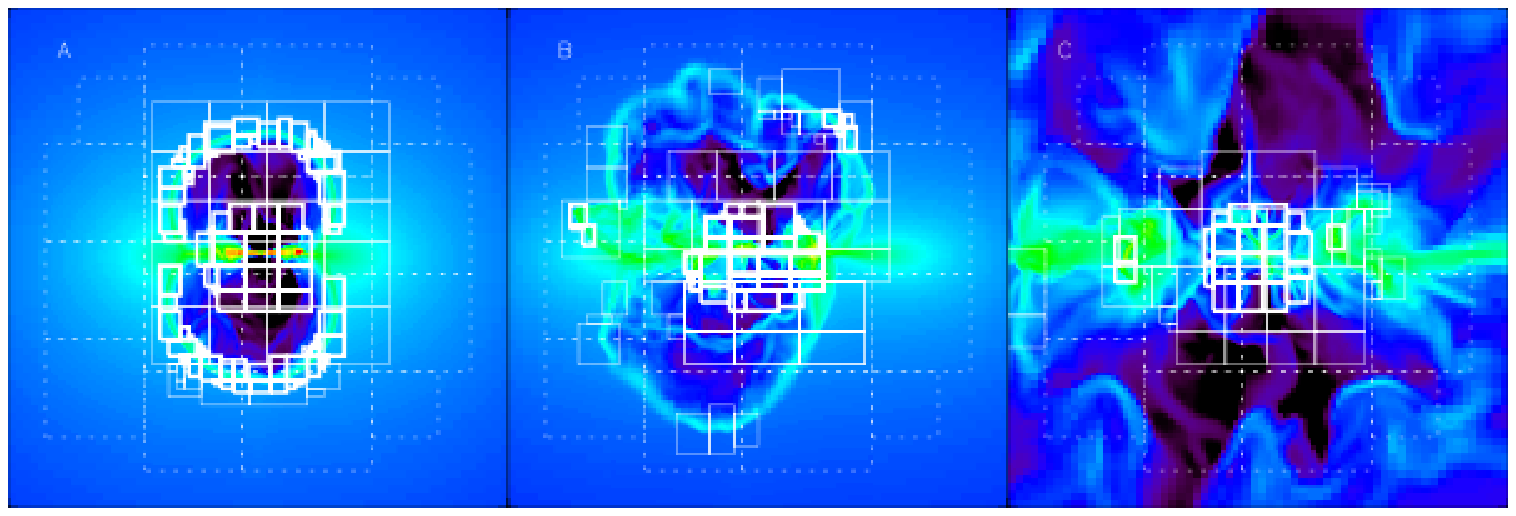}
\end{center}

\noindent {\bf Supp.\ Fig.\ 4.} Three snapshots of the simulation domain at the same times as those shown in Fig.\ 1C -- 1E in the main text. In this figure, the times are (A) 34.0 kyr, (B) 41.7 kyr, and (C) 55.9 kyr. Each panel shows a slice of density in the same plane and using the same color scale as in Fig.\ S1. The box shown is $(5000\mbox{ AU})^2$ in size. The overlayed rectangles show the locations of boundaries of our AMR grids where they cross the plane shown by the snapshots. Thick lines show level 6, the finest resolution level. Thin unbroken lines show level 5, and thin dashed lines show level 4.

\clearpage

\begin{center}
\includegraphics{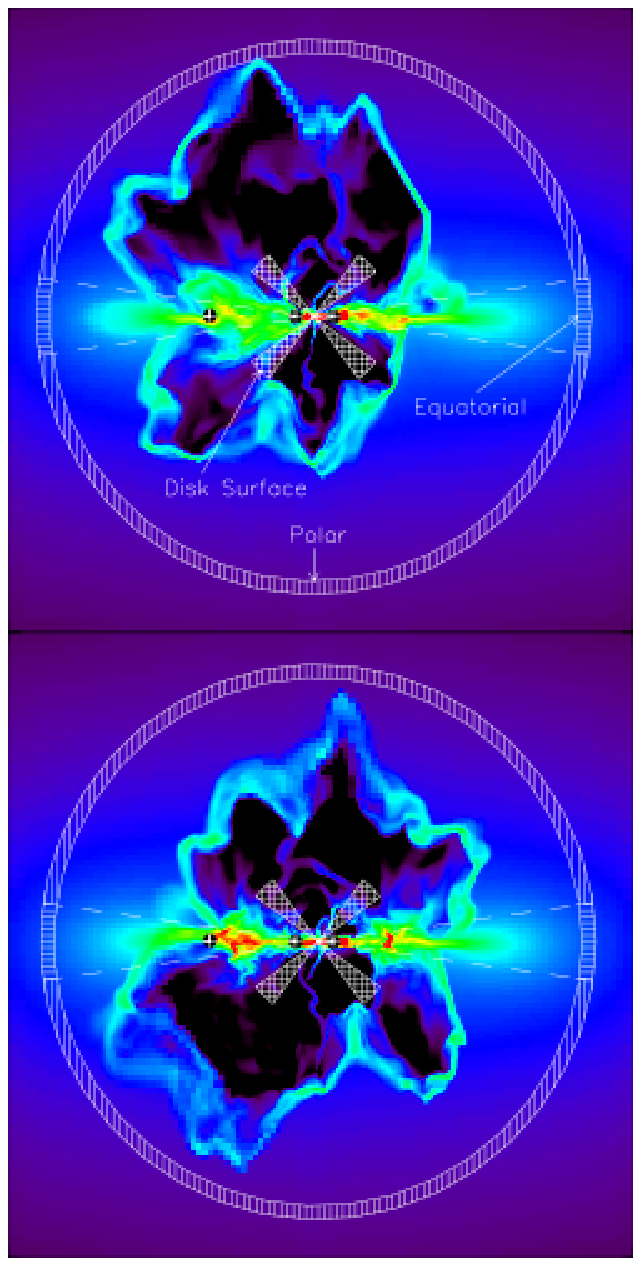}
\end{center}

\noindent {\bf Supp.\ Fig.\ 5.} Snapshots of density in two different slices through the simulation domain at 51.1 kyr of evolution. Color indicates density from $10^{-18}-10^{-14}$ g cm$^{-3}$ on a logarithmic scale. Plus signs show projected stellar positions. The upper and lower panels show two orthogonal planes that intersect at the rotation axis; the region shown in both cases is $(9000\mbox{ AU})^2$. The three hatched regions show the surfaces through which we compute the polar mass flux $\dot{M}_{\rm po}$ (\textit{vertical hatch}), equatorial mass flux $\dot{M}_{\rm eq}$ (\textit{horizontal hatch}), and disk surface mass flux $\dot{M}_{\rm ds}$ (\textit{cross hatch}), and the dashed lines show an angle $\alpha_e$ above and below the equatorial plane. The regions shown correspond to $r_1=3900$ AU, $r_2=1100$ AU, and $\alpha_e=8^{\circ}$.

\clearpage

\noindent {\bf Supp.\ Movie.\ 1.} An animation of the full simulation. The left column column shows column density projected onto the equatorial plane, and the right column shows volume density in a slice perpendicular to the equatorial plane. The top two frames show a $(0.25\mbox{ pc})^2$ region, and the bottom two show a $(4000\mbox{ AU})^2$ region. The color scales are logarithmic. Starting from the upper left image and proceeding clockwise, the ranges used for the color scale are $10^{-2}-10^2$ g cm$^{-2}$, $10^{-19}-10^{-15}$ g cm$^{-3}$, $10^{-19}-10^{-13}$ g cm$^{-3}$, and $10^{0}-10^{2.5}$ g cm$^{-2}$.

\clearpage


\begin{scilastnote}
\item 
We acknowledge support from: the NSF through grants AST-0807739 (MRK) and AST-0606831 (RIK, CFM); the Spitzer Space Telescope Theoretical Research Program, provided by NASA through a contract issued by the Jet Propulsion Laboratory (MRK); NASA through ATFP grants NAG 05-12042 and NNG 06-GH96G (RIK, CFM); and the US DOE at LLNL under contract B-542762 (RIK, SSRO, AJC).  This research used the Datastar system at the NSF San Diego Supercomputer Center (grant UCB267).
\end{scilastnote}

\end{document}